\documentclass[a4paper,twoside]{article}

    \usepackage{amsmath,amssymb}
    \usepackage{bm}
    \numberwithin{equation}{section}

\begin{document}

    \setlength{\baselineskip}{5mm}

\begin{center}

\vspace*{20mm}

{\LARGE \textbf{Spacetime in the Ultimate Theory}}

\vspace*{10mm}
{\large Noboru Nakanishi} \footnote{12-20 Asahigaoka-cho, Hirakata 573-0026,
Japan. E-mail: nbr-nak@trio.plala.or.jp}

\textit{Professor Emeritus of Kyoto University}

\end{center}

\vspace*{15mm}

If we assume that there is the ultimate theory at all,
how should the concept of the spacetime be formulated there?
The following essay is my consideration on such a question.
The use of mathematical expressions is suppressed as long as
possible. Any criticism on my opinion is welcome.  

\vspace*{10mm}

{\large \S 1. Motivation}\\

The universe consists of its container, \textit{spacetime}, and
its contents, \textit{substance}. The substance which we experience
daily is very much different from its intrinsic constituents,
\textit{elementary particles}. On the contrary, the spacetime
at the microscopic level is not significantly different from
the spacetime which we experience daily. Of course, the relativity
theory clarified that space and time must be treated not separately 
but in a unified way as a 4-dimensional manifold, but it did not
suggest any essentially new aspect in the microscopic structure
of the spacetime.

Historically, in Heisenberg's quantum mechanics, the
coordinates of the configuration space was formulated not as real
numbers (c-numbers) but as operators (q-numbers). However,
in Schr\"{o}dinger's wave mechanics, describing the
same contents, the spatial coordinates are formulated as
c-numbers by representing their conjugate momenta by 
differntial operators. The concept of the c-number spacetime
was carried over to quantum field theory by promoting the
wave functions to field operators. But quantum field theory
is most naturally formulated in the Heisenberg
picture but not in the Schr\"{o}dinger picture.
The most fundamental quantities in quantum field theory are
quantum fields, which are the operator-valued (generalized) 
functions of the c-number spacetime. 

The present standard theory of elementary particles is completely
describable by quantum field theory in the 4-dimensional spacetime.
Within this framework, there is no necessity for requiring any change
for the concept of the c-number 4-dimensional spacetime. Nevertheless,
most of particle physicists believe that this notion of the ``c-number
4-dimensional spacetime" should be extended in some sense.
The main reasons for this belief may be as follows.
 
\vspace*{3mm}
[A]  Because the standard theory is renormalizable, the physical S-matrix
can be written down so as to be free of divergences.
The theory, however, involves many ajustable parameters and 
the distribution of their actual magnitudes cannot be explained by the
theory at all. Hence, one hopes to resolve this difficulty
by extending the concept of the spacetime: Those parameters might become
calculable by making the unrenormalized quantities finite, or at least,
one could naturally understand the hierarchy of the actual values of those 
parameters.

\vspace*{3mm}
[B]  The action for quantum field theory is constructed so as to satisfy
the requirement of various symmetries. Thus the requirement of symmetries
is the principle governing the construction of the theory; therefore, 
the theory having larger symmetries would be regarded as more favorable
as the ultimate theory. Since the \textit{raison d'\^{e}tre} of internal symmetries
is not clear, it is more preferable that there are some intrinsic 
connection with the spacetime symmetry. Unfortunately, in the ordinary
4-dimensional spacetime, admissible spacetime symmetries are quite limited;
hence one wishes to find the large symmetry unifying both 4-dimensional
spacetime symmetry and internal symmetries by extending the concept of the
spacetime itself.

\vspace*{3mm}

[C]  Owing to the coherence of theories, the gravitational field, which 
was the spacetime metric in classical gravity, must be
quantized. Hence, in quantum gravity, the spacetime metric becomes
a q-number. Of course, this fact is different from the proposition that
the spacetime itself becomes non-commutative, but it means that the 
concept of ``two events separated by less than the Planck length"
cannot be well-defined at least in the geometrical sense.
Therefore, the ordinary structure of the conventional 4-dimensional
spacetime is not naively extendable to the Planck-scale world.

\vspace*{3mm}

The above three problems may be mutually related, but it will be too
fantastic to believe that they can simultaneously be resolved to achieve the 
ultimate theory, just as the superstring people dream.
It is more reasonable to attack each problem separately by
focusing on particular points. In what follows, I will not discuss
the superstring theory, as I criticized it elsewhere.\\

{\large \S 2. Discrete spacetime}\\

As is well known, the ultraviolet-divergence difficulty arises from the
behavior of the theory in
infinitesimally small spacetime distances. Therefore, it is natural to
suppose that if the spacetime is discrete, then the theory will be utterly
free of ultraviolet divergences. Various kinds of discrete-spacetime models
have been proposed so far.
But none of them has brought any promising achievement.

The essential difficulty of the discrete spacetime is how to formulate
the structure of the spacetime. Without using continuity, it is impossible
even to define the dimensionality of the spacetime as an invariant under
the point transformations.   
It is almost inevitable to introduce a background continuum
in order to formulate the theory by which quantitative calculations are
possible. Once, however, one introduces a background continuum, 
it immediately becomes the
\textit{fundamental spacetime}, because everything quantitative
must be defined in reference to it.

Thus, although the idea of the discrete spacetime is quite attractive, 
it is always destined to falling into a theory of the continuous
spacetime.\\
 
{\large \S 3. Higher-dimensional spacetime}\\

It is the cheapest idea to extend the spacetime to
the one having a dimensionality higher than 4.
Nevertheless, recently, it has become quite fashionable to investigate
the theory of extra dimensions. The grounds for considering it 
seem to be the excuse that any possibility is worth investigating
as far as it is not completely denied by experimental evidences
and the expectation that it might be possible to resolve 
the hierarchy problem stated in [A] by introducing new parameters 
in the extra-dimension space. It is quite disgusting to
revive the old  Kaluza-Klein theory without any essentially new
idea for resolving the fundamental problem of how to expel the
extra dimensions from the physical world. Furthermore, the higher-dimensional
theory makes ultraviolet-divergence difficulty out of control;
the extra-dimension people merely hope, without showing 
its ground, that the divergence problem
might be resolved by a non-perturbative treatment.

The dimensionality of the spacetime where we live is undoubtedly 4.
There is no other observation more manifest than this fact.
The extra dimensions, whose number is denoted by $N$, are qualitatively
different from the physical spacetime dimensions from the outset.
Recently, the superstring people have proposed the hypothesis that all
fields other than the gravity are confined to a soliton-like
(according to their claim) object called the ``$D$-brane", but it would be
quite difficult to formulate such an idea within the framework
of the $(4+N)$-dimensional quantum field theory.
A theory is worth being called a $(4+N)$-dimensional one if and
only if its fundamental action is $(4+N)$-dimensionally 
symmetric. But such a theory remains $(4+N)$-dimensionally
symmetric unless one breaks it artifically. In order to make
the $N$-dimensional extra space invisible, the extra-dimension people are
forced to make it round into a tiny one \textit{by hand}. This procedure implies
that they are actually considering the 4-dimensional spacetime 
accompanied with an $N$-dimensional \textit{internal} space. If so, they should 
honestly claim that the extra dimensions are internal.
Then there is no logical basis for adopting a $(4+N)$-dimensionally
symmetric action. It is quite non-scientific to pretend as
if such an action were a privileged one according to the symmetry
principle. 

There may be the objection that the extra dimensions are made
round not by hand but ``spontaneously". Certainly, there is no ``proof"
of the no-go theorem stating that the desired spontaneous compactification
can never take place. But if they assert the possibility of such
compactification, the responsibility of verifying it must be attributed 
to those who assert it. That is, the extra-dimension people
should construct at least one model in which the spontaneous
compactification of the extra dimensions takes place in a natural way.
I cannot believe that such a model can be constructed without greatly changing 
the framework of the conventional quantum field theory, because 
the situation encountered here is qualitatively different from the
ordinary spontaneous breakdown of symmetry. Since the dimensionality
is discrete, it is quite unreasonable to characterize the spontaneous
breakdown by such a continuous parameter as energy. But the most
difficult problem to be shown is \textit{not} how to compactify the extra
dimensions \textit{but} how to realize a \textit{trivial
product} of the physical spacetime and the compactified space.
That is, they must show the reason why the structure of the compactified
space does not vary point by point in the physical spacetime.
The extra-dimension people might obtain some favorable results
in the hierarchy problem, but they should recognize that
the cost of their achievement is deseparately high.

It is quite questionable to change the spacetime structure for the 
purpose of resolving such a quantitative problem as the hierarchy problem.
I believe that the introduction of a higher-dimensional
spacetime is too simple-minded to be on the right way to the
correct theory.\\

{\large \S 4. Supersymmetry}\\

If a higher-dimensional spacetime is introduced straightforwardly,
it becomes almost inevitably necessary to
discriminate the extra dimensions by hand. If so, it is more 
preferable to consider the possibility that such discrimination
exists in the built-in form. For example, one may suppose that
the extra dimensions are not real numbers but Grassmann numbers.
This possibility is realized as the supersymmetric theories;
in particular, SUSY is the theory obtained by supersymmetrizing the Poincar\'{e}
symmetry. It is the most beautiful theory which unifies the spacetime
symmetry and internal symmetries as is discussed in [B].
Unfortunately, however, it seems that NATURE does not adopt SUSY
as the fundamental principle. The reasons for believing so are as follows.

\vspace*{3mm}

(1)  Because SUSY is the unique symmetry which nontrivially includes the 
Poincar\'{e} symmetry, it has been believed that SUSY would be a privileged 
symmetry. But the uniqueness is a statement concerning the symmetry for
the physical S-matrix but \textit{not the symmetry for the fundamental action},
which is the most important object for constructing the theory.
Furthermore, evidently SUSY is not realized in the actual physical S-matrix.
Thus, there is no reason for believing that SUSY is a privileged symmetry. 
Later I will discuss this point in more detail.

\vspace*{3mm}

(2)  The experimental evidences clearly show that the actual mass spectra of
elementary particles are not consistent with the unbroken SUSY. Hence if 
SUSY is assumed to be a true symmetry, it must be spontaneously broken.
Then there must exist the Nambu-Goldstone fermion, but its existence is
completely denied experimentally. I am not certain whether or not
the ``super-Higgs mechanism", by which the Nambu-Goldstone fermion
is absorbed into the gravitino in the framework of supergravity,
can be well formulated. But even if it can be well formulated, such explanation 
for the absence of the Nambu-Goldstone fermion implies that
the conventional approximation  of neglecting gravity in particle physics
is violated already in the low-energy region.
Therefore, if one insists on the super-Higgs mechanism, 
such a standpoint implies to assert that
the great success made so far in particle physics is no more than
mere accidental luck, because one has admitted the possibility that
quantum gravity may have significant
contributions in the low-energy region. 

\vspace*{3mm}

(3)  In spite of the fact that so many elementary particles have been 
found experimentally, utterly no superpartner-like particles predicted by
SUSY are discovered. It is quite unnatural to assume that all 
superpartners have a mass so large that no present-day accelerators
can produce them \textit{without exception}. I believe that NATURE does not
plan a ``perfect crime". If SUSY is really a physical symmetry,
something should indicate its existence in the low-energy region,
just as the muon demonstrated the existence of the second generation
in the early stage of particle physics.

\vspace*{3mm}

If NATURE does not adopt SUSY, supergravity and superstring 
must be abandoned, and many people researching them will become greatly embarrassed.
But, of course, their hope on the validity of SUSY does not
guarantee its actual existence.
I believe that scientists must always be humble for NATURE.
People should not so much adhere to SUSY but investigate other
possibilities. 

As is well known, in
 quantum theories of gauge fields and the gravitational field, one
defines the physical subspace by setting up the Kugo-Ojima subsidiary
condition in order to avoid the appearance of negative-norm states.
That is, the physical states are defined as the states which are 
annihilated by the BRS generator. Then the physical S-matrix,
namely, the S-submatrix between the physical states becomes unitary
(\textit{modulo} zero-norm states). Accordingly, when the states predicted by 
a particular symmetry are unphysical, they do not appear in the physical
S-matrix even if that symmetry is unbroken. 
In the framework of indefinite-metric quantum field theory, therefore,
many unbroken symmetries are available  without predicting the existence 
of extra physical particles, so that one can construct the action
having a large (super)symmetry which may unify the spacetime symmetry
and internal symmetries without contradicting the no-go theorem for the
extension of the Poincar\'{e} symmetry other than SUSY.\\

{\large \S 5. Quantized spacetime}\\

As pointed out in [C], two events separated by less than
the Planck scale cannot be well defined in quantum gravity.
Accordingly, one must not expect that the geometrical structure of
the ordinary spacetime remains to exist in the region less than
the Planck scale. This fact motivated for many researchers to consider
the quantization of the spacetime.

Thinking carefully, however, this logic has a big jump. While what quantum 
gravity claims is the proposition that the ordinary geometrical relation 
loses its meaning for \textit{two} (very near) spacetime points, the quantization of 
the spacetime asserts that nontrivial commutation relations are set up
between the components of \textit{only one} spacetime point.
There is neither logical nor physical ground for introducing such non-commutativity 
between the components of a spacetime point. The quantized spacetime is
totally irrelevant to quantum gravity and the uncertainty between the 
components of a spacetime point has no inevitable connection with the
Planck length. If such connection should have been derived, it would 
be a consequence of an assumption introduced tacitly.

Indeed, when Snyder proposed a theory of quantizing the spacetime half
a century ago, he did not care about quantum gravity at all. 
His purpose was to resolve the divergence problem of quantum field
theory. Recently, the quantized spacetime has been revived owing to the 
fashion of the non-commutative geometry. But the quantized spacetime
seems to be investigated in favor of the mathematical
interest rather than the physical requirement. 

In most cases, what is called the quantized-spacetime
quantum field theory
is what is obtained by merely transcribing $x^\mu$ from c-numbers
to q-numbers but keeping the framework of the conventional quantum field theory  
unchanged. The commutation relations for $x^\mu$ are set up artificially
without any logical connection with the fundamental principle of quantum
field theory. Indeed, in spite of the fact that,
except for the 2-dimensional case, the commutator $[x^\mu,x^\nu]$
cannot be defined Lorentz covariantly as far as 
it is a c-number, what is adopted as the action
is ``what becomes Lorentz invariant if the non-commutative $x^\mu$ is regarded as
commutative". Once the Lorentz invariance is violated in the
commutator, the theory is no longer Lorentz invariant so that
the action may be Lorentz non-invariant in the same order of magnitude. 
Nevertheless, the quantized-spacetime people do not consider
the possibilities of intrinsically non-invariant actions at all.
 I believe that the genuine quantized-spacetime
theory, if any, should have some organic relationship between the 
action and the q-number property of $x^\mu$, and it should be derived
from the fundamental principle in a unified way.

In considering the quantized-spacetime theory constructed
artificially as stated above, one must be very careful about 
identifying the quantity which has happened to be written  $x^\mu$
with the quantized spacetime. In general, there are infinitely
many q-number quantities which correspond to the macroscopic
spacetime where we live. There is no reasonable criterion to select
the genuine quantized spacetime from them in such an artificially
constructed theory. If there is a c-number candidate among them
(for instance, if $[p_\mu, p_\nu]=0$,
where $p_\mu$ denotes the momentum canonically conjugate to 
$x^\mu$, and if
``$x^\mu$ minus a certain function of $p_\nu$" is commutative
with anything),
it is meaningless, at least for the purpose of 
constructing a fundamental theory, to adopt a non-commutative one as $x^\mu$;
it is no more than the quantized-spacetime people's
wishful thinking.

Suppose that the system of $x^\mu$ and $p_\nu$ is algebraically
closed. If one does not wish to encounter the operator-ordering 
problem on the right-hand sides of commutation relations,
the commutators $[x^\mu,x^\nu]$, $[x^\mu,p_\nu]$, and $[p_\mu,p_\nu]$
should be linear with respect to $x^\mu$ and $p_\nu$ 
(Furthermore, dependence on $x^\mu$ is forbidden by translational 
invariance.). In order to make it possible to have
a more general quantized spacetime, therefore,
it is preferable to introduce a new set of c-number parameters
and express  $x^\mu$ and $p_\nu$ in terms of those parameters and 
their differential operators.
In this formulation, however, the space defined by those
parameters plays the role of the fundamental spacetime 
rather than a mere mathematical tool. That is,
such a theory should be regarded as
a kind of the higher-dimensional-spacetime model,
though it is unnecessary
to make round the extra dimensions by hand
in contrast to the model discussed in \S 2. 
Instead, the real problem to be clarified is on what principle
the expression for $x^\mu$ in terms of the parameters
is derived. I cannot imagine that it is derivable by a
kind of the action principle.

I emphasize that the quantized spacetime has neither logical ground
for its indispensability nor the principle on which it is based.
Hence the calculations concerning it often become no
more than mathematical exercises.
Of course, one may claim that the quantized spacetime is
an approximation of string theory, but such an assertion implies that
the quantized-spacetime theory itself is not a candidate of the 
ultimate theory. \\

{\large \S 6. Geometrical structure}\\

General relativity reduced gravity to geometry.
Also in quantum theory, many geometrical concepts, such as
solitons, monopoles, instantons, etc., have been introduced.
Particularly, in quantum field theory, 
the notions of topological geometry have been regarded as
important in connection with the problem of anomaly.
Recently, this tendency has become more and more intense,
keeping pace with the fashion of the path-integral approach.
This is because, since the path-integral formalism
directly gives the Green functions from a c-number
action, classical notions can easily be incorporated into the framework 
of quantized theory.   

Although certainly the path-integral approach is very convenient
for calculations, one should not forget the fact that the
unitarity of the physical S-matrix cannot be proved without
the help of the operator formalism. While only the \textit{variation} of 
the action is relevant (according to the action principle)
in the operator formalism, the \textit{action itself} 
is regarded as a meaningful quantity in the path-integral formalism.
Indeed, the value of the action is often connected with a
topological invariant. It should be re-examined more critically, 
without believing its relevance \textit{a priori},
whether or not such a quantity is really physically meaningful.

If the spacetime coodinates are written $x^\mu$ universally over the 
\textit{whole} spacetime, one is necessarily considering a
topologically trivial manifold. In order to describe a
more general manifold, it is necessary to decompose it into patches,
and after introducing the coordinate system to each patch, 
one should join together those patches consistently.
The quantity $x^\mu$ is merely the name of a point on the manifold,
and it varies patch by patch. Therefore, if the theory is formulated
by universally using $x^\mu$ as the spacetime coordinates,
most people regard such a formulation as a method devoid of 
generality.

This way of thinking is correct if one is discussing a model of
quantum mechanics. In this case, the manifold is \textit{given from the 
outset}, and it is usually connived why a topologically
nontrivial manifold has become relevant to the
problem; maybe it is what is set up by experimentalists.
I believe, however, that \textit{the same is no longer true in quantum
field theory}. There, $x^\mu$ is a microscopic quantity involved only in the 
quantum fields but not the macroscopic spacetime which we directly experience.
The quantum-field-theoretical structure of the spacetime is governed
by the spacetime symmetry of the action, such as the Poincar\'{e} 
symmetry. It is the fundamental symmetry as a physical principle,
which can never be modified by the experimentalist's will.
I believe that the fundamental theory should not start with 
the following proposition: ``There existed a manifold at the beginning".

Some of the cosmology people claim that, because the big-bang expanding
universe is the unique actual spacetime, it is more adequate to 
quantize the theory in the expanding universe rather than to do 
in the Minkowski space. If they insist on such a standpoint, however,
they should not \textit{a priori} postulate that the universe
is homogeneous and isotropic, but solve the Einstein equation
rigorously by substituting the actual distribution of the matter
into its right-hand side. Of course, it is impossible to 
work out this task. In general, it is unreasonable to formulate the fundamental
theory of physics by taking the historically accidental facts
as the premise for constructing it.  

When one wishes to formulate quantum gravity geometrically by means of the
path-integral formalism, there arises such a notion as the ``sum
over all possible manifolds". But this notion is quite ambiguous.
Furthermore, it is not clear whether or not the metric signature,
such as Lorentzian or Euclidean, is prescribed beforehand.
If the metric signature is fixed to a particular one from the outset, it
contradicts with the conventional definition of the functional integral over
the metric field $g_{\mu \nu} (x)$.
Of course, a manifold of Lorentzian metric cannot be obtained from
that of Euclidean metric by the so-called ``Wick rotation".
I think that there is no logical basis for the expectation that
quantum gravity can be constructed by means of the ``sum over all 
possible manifolds".

Such a geometrical notion as the manifold can be well-defined
only if the neighborhoods of each point are set up.
The concept of the ``neighborhood system" is \textit{classical}.
Indeed, it is impossible to introduce such a topological concept as
the neighborhood consistently logically prior to the 
determination of the metric signature of the manifold.
That is, in a topologically well-defined manifold, the relation  
between any two points is always definite, but not uncertain as 
is suggested by quantum gravity. It is quite questionable,
therefore, whether or not the spacetime structure implied by quantum 
gravity can be reproduced by summing up the exponentiated action
over such topologically well-defined manifolds \textit{\`{a} la}
Feynman. It seems to me that one should include in the sum  
some curious objects which may not necessarily constitute manifolds globally.   

It seems quite unnatural in the ultimate theory to adopt either
a particular manifold or all possible manifolds as the spacetime 
structure.\\

{\large \S 7. Spacetime structure in the ultimate theory}\\

Human being can observe and recognize only \textit{classical} objects.
However, almost all physicists believe that the fundamental
law of physics governing NATURE is of \textit{quantum} theory. NATURE always 
talks in the language of quantum theory, but in order to understand
it, human being must translate it into the language of classical theory
by introducing a foreign notion, ``probabilistic interpretation", 
for the observation of quantum systems.  
Therefore, we do not wish to incorporate the theory of the observation
into the framework of the ultimate theory. Thus, apart from the theory of
the observation, it is quite natural to expect that quantum theory 
is \textit{closed} within itself without aid of 
classical theory, because it is impossible to believe that
NATURE has utilized the notion 
invented by human being in order to control itself. 

Thus, I want to propose the following fundamental principle of \textbf{``quantum priority"}:

\vspace*{2mm}
\hspace*{1mm} {\large \textit{In the ultimate theory, any concept of classical 
physics must not}

\hspace*{1mm} \textit{appear logically prior to its quantum-theoretical 
construction.}}

\vspace*{2mm}

The ground for proposing the above principle
is as follows: If there exists a
classical quantity appearing logically prior to the quantum-theoretical
construction, such a quantity is necessarily brought into the theory from its 
outside; this fact means that the theory is not closed within itself.
Accordingly, it cannot be regarded as the ultimate theory.
Since the pricinple of quantum priority is quite natural, there may be few people
who positively object it. Nevertheless, if it is applied to the
discussion of the spacetime structure, it yields a very powerful
restriction on the ultimate theory.

According to the above principle, \textit{any manifold cannot be identified with 
the spacetime}.  This is because to choose a particular manifold 
is justifiable only by some classical physics, as long as
one does not introduce it by hand.
This remark is applied even to the Minkowski space, because the
existence of the light cone is evidently a consequence of a particular 
classical theory, \textit{special relativity}! Thus, the ultimate theory 
rejects even the microcausality, which is one of the axioms adopted by the
axiomatic quantum field theory.

Then there arises a question: Does it contradict the above principle
to consider $x^\mu$ as the variables of quantum fields? No! 
This is because $x^\mu$
merely represents  a set of 4 real numbers and the concept of the
real-number field \footnote{The word ``field" is used in the sense of \textit{K\"{o}rper}.}
 can be defined by the completion of the rational-number field
 without employing the notion of the neighborhood.
Of course, there may be the objection that the Minkowski
space is also a concept definable purely mathematically.
But, while evidently the Minkowski space was introduced under the
special reference to special relativity, the real-number
field is a generic concept which can be used universally without
restricting its physical applications.
That is, the real-number field is not what arises as 
a consequence of any particular classical physics.
Thus I believe that the use of real numbers does not contradict
the principle of quantum priority.

The real-number field has a very special element ``0".
In order to guarantee that physical principles do not 
depend on the special property of 0, it is sufficient for the
theory to be invariant under the change of $x^\mu$ by a constant
$\alpha^\mu$. Therefore, it is quite natural to require the theory
to be translationally invariant. Furthermore, it is preferable
that the theory is constructed in such a way that
the 4 real numbers $x^\mu$ ($\mu =0,1,2,3$) be not completely
mutually unrelated but all linear combinations of them have the 
equal right. That is, it is natural to require general linear invariance.
The combined symmetry of translational invariance and general
linear invariance is called the \textit{affine} symmetry. The 
(4-dimensional) affine transformation 
is a respectable transformation, because it is the unique
analytic one-to-one mapping 
from $\mathbf{R}^4$ to $\mathbf{R}^4$.
The affine symmetry may be regarded, therefore, as the most natural    
symmetry, as long as the theory is based on $\mathbf{R}^4$.
The affine geometry was born purely from the mathematical interest,
that is, it is totally irrelevant to any classical theory of physics.
If the ultimate theory is constructed in terms of quantum fields, 
it is most adequate to require the affine symmetry.

There may still arise some questions: Why are 4 real numbers
$x^\mu$ ($\mu =0,1,2,3$) considered? i.e., why no more than 4? and why
not complex numbers?  That is, the criticism states that to consider 
4 real numbers might be a consequence of classical physics.
But I think that the fact that any event is assigned by 4 real numbers
is the most common experience rather than classical physics.
As long as we do not admit to make any by-hand procedure in the 
process of relating $x^\mu$ to the macroscopic spacetime,
it is extremely unplausible that under any other assumption one 
succeeds in deriving the actual macroscopic spacetime.\\

{\large \S 8. Quantum gravity and the spacetime structure}\\

In \S 7, I have discussed that, as far as one considers
the quantum-field-theoretical formulation of the ultimate theory,
it is most natural to require the affine symmetry for $x^\mu$.
When formulating quantum gravity, the affine symmetry can be incorporated
 quite naturally. As is well known, although the classical Einstein
gravity is invariant under general coordinate transformations,
it is impossible to quantize the gravitational field without
breaking such  a local symmetry, that is, it is necessary to introduce 
a gauge-fixing term and a Faddeev-Popov ghost term in such a
way that their sum becomes BRS invariant. After doing this, 
the largest symmetry, directly related to $x^\mu$, which can remain unbroken
is nothing but the affine symmetry. Indeed, as long as 
such a particular classical metric as the Minkowski metric is not brought
into the gauge-fixing term, the affine symmetry should survive in the
quantum Einstein gravity.

Many people seem to assert that the gauge-fixing term is no 
more than a necessary evil because the physical S-matrix 
does not depend on its choice. But I point out that this assertion is
based on the standpoint of giving the priority to classical theory.
Quantum gravity starts with a BRS-invariant action; if the gauge-fixing
terms are different, then the corresponding theories are different 
\textit{as quantum theories}.
In the classical electromagnetism, the observable quantity is the
field strength $F_{\mu \nu}$ and therefore the gauge freedom of
the potential $A_\mu$ is physically meaningless, but 
$A_\mu$ is the quantity fundamentally important in quantum theory.
Indeed, such an observable phenomenon as the Aharanov-Bohm
effect cannot be explained without taking $A_\mu$ into account.
Although no example of the observable effect which cannot be
explained without taking the gauge-fixing term into account is found as yet,  
 the existence of the critical dimension of the string theory
was found to be \textit{not} irrelevant to the choice of the gauge-fixing term in
the 2-dimensional quantum gravity.
At any rate, hereafter, I assume that there exists \textit{uniquely}
the right choice of the gauge-fixing term according to the principle of
quantum priority.

The de Donder gauge fixing, which may be regarded as the most natural
one, can be realized if one employs the B-field formalism,
where the B field $b_\mu$ is an auxiliary bosonic field appearing 
in the gauge-fixing term only. 
The BRS-invariant theory can be constructed in the most transparent
way by introducing the notion of the ``intrinsic BRS transformation",
which is the conventional BRS transformation minus its orbital part.
Then carrying out the canonical quantization, one can explicitly
calculate all equal-time (anti-)commutation relations in closed form.
The quantum theory of gravity thus constructed turns out to have a 16-dimensional
supersymmetry 
\footnote{More precisely, its superalgebra
 is the (8+8)-dimensional inhomogeneous orthosymplectic
superalgebra consisting of 144 generators. Of course, 
the affine algebra is its subalgebra.}
 based on the ``16-dimensional supercoordinates"  
$\{ x^\mu, b_\nu, c^\lambda, \overline{c}_\rho \}$, 
where $c^\lambda$ and $\overline{c}_\rho$ denote
the Faddeev-Popov ghost and anti-ghost, respectively; the outstanding
beauty of this theory may be regarded as a support for the assertion that 
the de Donder gauge fixing is of the right choice. 
There may be the objection that in the canonical quantization 
the \textit{time} is treated in a specially distinguished way, but such a criticism
does not apply to the theory having the affine symmetry.
This is because, in the affine-invariant theory, any linear combination
of $x^\mu$ can be used as the ``time for the canonical quantization"
and the results are essentially independent of its choice.

Although the Einstein equation is written in terms of rational functions
of $g_{\mu \nu}$ (and its derivatives), the action integral and the
symmetry generators need the square root of $- \mathrm{det} g_{\mu \nu}$.
Therefore, its hermiticity is not guaranteed unless certain restriction
is imposed on $g_{\mu \nu}$. If one avoids to introduce artificial
restriction, the following procedure is the unique natural resolution: 
The fundamental field is  the ``vierbein" field
$h_\mu \, ^a$ but not $g_{\mu \nu}$, which is expressed as
$g_{\mu \nu} = \eta_{ab} h_\mu \, ^a h_\nu \, ^b$.
Here, the ``internal" Minkowski metric $\eta_{ab}$ is 
a quantity totally irrelevant to $x^\mu$ at this stage.
I emphasize that \textit{the introduction of $\eta_{ab}$ does not imply that of 
the Minkowski space; it merely assigns the Lorentzian signatrue
to  $g_{\mu \nu}$.}
Correspondingly, one should forget about the classical interpretation,
``coordinate system of the tangent space", of the vierbein field.

The introduction of the vierbein field has brought 6 extra degrees
of freedom corresponding to the local internal Lorentz transformations,
but they must not bring any new physically 
observable effects. To guarantee this, one introduces the gauge-fixing
term for the local internal Lorentz invariance and the corresponding
Faddeev-Popov ghost term so as to become (local-Lorentz) BRS-invariant.
Here, it is very important to choose the gauge-fixing term 
so as to keep the general coordinate invariance
and the global internal Lorentz invariance unbroken.

It is undoubtedly true that there exist Dirac fields (more
precisely, Weyl fields, according to the electroweak theory) as
the elementary-particle fields. As is well known, the generally
covariantized Dirac field can be formulated by using the vierbein.
In this formulation, however, the spinorial transformation property
of the Dirac field is necessarily transferred to the internal
Lorentz freedom, and the Dirac field becomes a spacetime scalar.
It is mathematically impossible to have a spacetime spinor field 
as far as one considers it in the framework of classical gravity.
As is explained in the following, however, this problem is satisfactorily
resolved in the framework of quantum gravity.

In quantum gravity, the global gauge symmetries remaining after gauge fixing 
are the affine symmetry and the global internal Lorentz invariance.
However, those symmetries are spontaneously broken.
The reason for this is that the veirbein field has a non-vanishing 
vacuum expectation value. If translational invariance is violated,
it essentially means that the natural law itself would depend on a
particular classical background. Since this is quite an unwelcome situation,
I assume that translational invariance is not spontaneously broken.
Then the vacuum expectation value,
$\langle 0|h_\mu \, ^a |0 \rangle$, of the vierbein field becomes
a \textit{constant} matrix independent of $x^\mu$. It must be a non-singular
matrix (i.e., its determinant is non-vanishing)
so as to make the spacetime non-degenerate. This is because, since 
$\langle 0|g_{\mu \nu}|0 \rangle$ gives the spacetime metric, 
its determinant must be non-vanishing if the spacetime is non-degenerate; 
hence the determinant of $\langle 0|h_\mu \, ^a |0 \rangle$
must be also non-vanishing (at least generically).
From this fact, one can easily show that all degrees of freedom of
general linear invariance and internal Lorentz invariance
are spontaneously broken.

Here, however, it is extremely important to note that there still remains
a certain Lorentz symmetry which is not spontaneously broken.
The generators of this unbroken symmetry are certain linear combinations
of the antisymmetric-part generators of general linear transformations
and the generators of internal Lorentz transformations with
coefficients depending on $\langle 0|h_\mu \, ^a |0 \rangle$.
The \textit{physical} spacetime is \textit{defined} by transforming $x^\mu$
by the matrix $\langle 0|h_\mu \, ^a |0 \rangle$.
Then the above unbroken Lorentz 
symmetry is precisely the spacetime Lorentz symmetry of 
particle physics. Indeed, one can confirm,
by explicitly calculating the commutators between the Dirac
field and the generators of the above symmetry, that
the Dirac field transforms as a spacetime spinor under it.

Thus, both the physical spacetime and the Lorentz symmetry of 
particle physics are the \textit{secondary concepts} appearing 
as a consequence of the spontaneous breakdown of symmetries, just as
the electromagnetic $U(1)$ symmetry is in the electroweak theory.
The above consideration suggests that supergravity
cannot be a fundamental theory because it is formulated under the
assumption of regarding
the Lorentz symmetry as a fundamental one.

As an additional remark, it should be noted that the Nambu-Goldstone
boson corresponding to the symmetric-part generators of 
general linear transformations is nothing but the graviton.
This fact guarantees the exact masslessness of the graviton.\\

{\large \S 9. Criticism and counterargument}\\

I have shown that quantum gravity can be satisfactorily formulated
under the principle of quantum priority. According to my establishment,
the set of 4 real numbers $x^\mu$ yields the physical spacetime
as a consequence of the spontaneous breakdown of the general linear
invariance.

Unfortunately, however, there are many researchers who do not wish to   
regard the quantum Einstein gravity as (a part of) the ultimate theory.
Such a standpoint may be based on the following reasons, but I wish to
show that most of them are essentially groundless.

\vspace*{3mm}
(1)  There are some people who identify the covariant formulation
of the quantum Einstein gravity with the covariant perturbation theory of it;
since the latter is unrenormalizable, they claim that it is not a physically
sensible theory. This criticism is, however, merely due to their 
erroneous identification; indeed, 
what is not adequate is to apply the covariant perturbation theory
to quantum gravity. The interaction picture, on which the covariant
perturbation theory is based, can be introduced 
under the assumption that $g_{\mu \nu}$ can be written as a sum of
a c-number metric and a quantum gravitational field of order
$\sqrt{\kappa}$, where $\kappa$ denotes the Einstein gravitational constant;
that is, the covariant perturbation theory can be introduced under the assumption
that $g_{\mu \nu}$ tends to a c-number as $\kappa \to 0$.
However, this assumption is \textit{wrong}! In the BRS-formulated theory of
quantum gravity, $g_{\mu \nu}$ actually tends to a \textit{q-number}
as $\kappa \to 0$. Therefore, the covariant formalism of the quantum Einstein gravity
should be solved \textit{in the Heisenberg picture}. The method for doing this
has been developed, and the $\kappa \to 0$ limit of $g_{\mu \nu}$
was explicitly calculated. Of course, it is unclear as yet whether
or not the divergence problem can be resolved in the Heisenberg-picture
approach. But, at any rate, I believe that it is very unreasonable to
reject the quantum Einstein gravity in such an uncertain situation.

\vspace*{3mm}
(2)  In the quantum Einstein gravity, it is impossible to resolve the 
problems stated in [A] and [B], and therefore it is not the unified 
theory. But why should all problems be solved simultaneously?
I think that there is no reason why
 the problem [A] is inevitably connected with the problem of the
spacetime structure and quantum gravity.
As for the problem [B], for example, there is a possibility of supersymmetrizing
the local internal Lorentz symmetry \textit{only}.
This possibility was investigated in detail; the
conclusion of this supersymmetric theory
is that the admissible gauge symmetry of 
particle physics is the chiral $SO(N) \times SO(N)$. 

\vspace*{3mm}
(3)  Some of the cosmology people believe that quantum gravity is
meaningless if the beginning of the universe cannot be discussed by it.
But the problem of the beginning of the universe must include the quantum
observation theory in the case in which there is no \textit{external} observer.
Therefore, it is a problem totally out of control; and indeed they
themselves do not seriously discuss the problem of the probabilistic
interpretation of quantum theory in studying the early
universe. Although the cosmology people prefer to adopt geometrical
approach, I believe that the theory
of quantum gravity should be formulated so as to be mathematically coherent to the 
operator formalism of gauge theory, because the latter has been well
established in particle physics.

\vspace*{3mm}
(4)  Some of the superstring people believe that it is fruitless to consider
any other theory because the superstring theory is the unique
candidate of the ultimate theory. I believe, however, that such an
assertion is too much prejudiced, because there are several
fundamental difficulties in the
superstring theory and nothing of them have yet been resolved.
I will not discuss them here, but I wish to point out  
that there are absolutely neither theoretical nor experimental 
evidences which justify the huge extension of the theoretical framework 
done in the superstring theory.

\vspace*{3mm}
In conclusion, I would like to state the following aphorism:

\vspace*{3mm}

\hspace*{1mm} {\large \textit{Do not seriouly take geometrical and classical images into
account in} 

\hspace*{1mm} \textit{constructing the ultimate theory.}}

\vspace*{3mm}
Since the grammar of the language spoken by NATURE is mathematics, the mathematical
coherence should be the key to the ultimate theory. 

\vspace*{3mm}
I hope that  more researchers of the fundamental physics
recognize the possibility that quantum Einstein gravity may already provide the
natural framework of the ultimate theory.\\

{\small
\begin{center}
\textbf{References}\\
\end{center}

As for the quantum Einstein gravity, see

\hspace{5mm}
N. Nakanishi and I. Ojima, \textit{Covariant Operator
Formalism of Gauge Theories and Quantum} 

\hspace*{8mm} \textit{Gravity}, World Scientific (1990), Chap.5.

 As for the method of solving quantum field theory in the Heisenberg 
picture, see 

\hspace{5mm}
 N. Nakanishi, Prog. Theor. Phys. \textbf{111} (2004), 301. 

As for the supersymmetric extension of the local internal Lorentz
symmetry, see

\hspace{5mm}
 M. Abe, Int. J. Mod. Phys. \textbf{A5} (1990), 3277.

\end{document}